# Benchmarking real-time monitoring strategies for ethanol production from lignocellulosic biomass

Pau Cabaneros Lopez[1], Hannah Feldman[1], Miguel Mauricio-Iglesias[2], Helena Junicke[1], Jakob Kjøbsted Huusom[1], Krist V. Gernaey[1]

[1]*PROSYS Research Center, Department of Chemical and Biochemical Engineering, Technical University of Denmark (DTU), Building 229, 2800 Lyngby, Denmark {pacalo, hafe, heljun, jkh, kvg @kt.dtu.dk}*

[2]*Department of Chemical Engineering. Universidade de Santiago de Compostela. 15782, Santiago de Compostela, Spain. miguel.mauricio@usc.es*

Corresponding author: Krist V. Gernaey, kvg@kt.dtu.dk

**Abstract**

The goal of this paper is to review and critically assess different methods to monitor key process variables for ethanol production from lignocellulosic biomass. Because cellulose-based biofuels cannot yet compete with non-cellulosic biofuels, process control and optimization are of importance to lower the production costs. This study reviews different monitoring schemes, to indicate what the added value of real-time monitoring is for process control. Furthermore, a comparison is made on different monitoring techniques to measure the off-gas, the concentrations of dissolved components in the inlet to the process, the concentrations of dissolved components in the reactor, and the biomass concentration. Finally, soft sensor techniques and available models are discussed, to give an overview of modeling techniques that analyze data, with the aim of coupling the soft sensor predictions to the control and optimization of cellulose to ethanol fermentation. The paper ends with a discussion of future needs and developments.

**Keywords**

# 1 Introduction

The monitoring of bioprocesses in real-time is a widely studied area, as real-time measurements of reactor conditions allow for a higher degree of control and process optimization than off-line monitoring [1]. In large scale biotechnology processes there is usually only a rather limited capability for real-time monitoring of the process due to lack of suitable – and affordable – monitoring techniques. Monitoring applications have been developed mainly for laboratory use [2]. There are many reports on the availability, advantages, and challenges of different monitoring techniques, but large scale monitoring in real-time with advanced sensors is rarely done. This is because there are hardly any investigations on the potential benefits of these methods [3]. Kiviharju *et al.* [4] compared different monitoring methods based on specific requirements for biomass monitoring, providing a guide to select the appropriate method under specific conditions. A number of papers review specific monitoring devices, in which the devices are described as single entities. For instance, Marison *et al.* [5], gave an extensive review of near infrared spectroscopy (NIR), mid infrared spectroscopy (MIR), Raman spectroscopy, dielectric spectroscopy (DS), and biocalorimetry, and Marose *et al.* [6], studied in situ microscopy, NIR spectroscopy, and fluorescence spectroscopy. Nevertheless, as the performance of these methods for specific monitoring objectives was not compared, it did not provide the reader enough information to support the selection of a given alternative.

Cellulose-based biofuels are produced from biomass mainly consisting of plant material, in which sugars are fixed in structures of cellulose and hemicellulose that are intertwined with lignin [7]. The cost of cellulose-based biofuels production cannot yet compete with non-cellulosic biofuels [8], in which the carbon source comes from relatively simple and easily accessible sources such as corn or sugar beet **(FIGURE 1)**. Non-cellulosic biofuels have been produced for more than two decades [9] and are now a mature technology given the considerable experience gathered operating and building plants for non-cellulosic biofuel production. As a consequence, monitoring is essential in cellulosic biofuel production in order to ensure that the process runs at the optimal process conditions and to compensate the relative lack of process understanding of this technology [10]. One of the goals of monitoring cellulose to ethanol fermentation is to increase the profit associated with the process [11]. An increased profit can be obtained by reaching a high ethanol yield (income increase) and by running the process under non-sterile conditions (cost reduction). However, there is an increased risk of contamination when working under non-sterile conditions, which would decrease the ethanol yield, compared to a sterile process. Monitoring the process to detect contaminations is therefore of importance to be able to stop the process as soon as a contamination is detected and avoid the loss of substrate.

Another challenge in cellulosic ethanol production is overcoming the action of inhibitors. Indeed, inhibition decreases the productivity, which makes the process last longer, and thus increases the costs. Furthermore, the longer the process lasts, the higher the risk of contamination. Monitoring of inhibitory components in the feed and the reactor is therefore needed so that a real-time strategy can be applied to improve the process performance. As the introduction of novel control and optimization techniques has a cost related not only to the equipment and implementation but also to the training of operators, the benefits must be demonstrated and clearly outperform the current process as it is operated. In optimal conditions, one would desire direct measurements of all components of interest – for example, substrates, biomass, inhibitory substances, and product concentration. However, this is not an economically viable option, due to the high costs associated with installing and maintaining the equipment needed to establish the different monitoring techniques.

The contribution of this study is to assess the alternatives for real-time monitoring of fermentation and link them with industrial challenges faced during ethanol production from lignocellulosic biomass. The application of combined techniques for advanced monitoring is covered for the first time. Beyond the review made by Pohlschleidt *et al.,* [12], this study explicitly relates the monitoring equipment and combinations thereof with the specific objectives of the process, in particular for the production of cellulosic ethanol [12]. Furthermore, this study evaluates the potential benefits of the methods with a case study involving the production of cellulosic ethanol. This is a relevant case study, as the complex feed stream containing multiple carbon sources, inhibitors, and particulates needs accurate monitoring to obtain knowledge of the process characteristics **(FIGURE 1)**. Furthermore, because the feed stream contains a significant amount of solid particles, the monitoring techniques need to be able to distinguish between relevant and irrelevant compounds.

The organization of this paper is such that section 2 describes the process layout of a cellulosic ethanol fermentation. Section 3 focuses on the added value of monitoring different key process variables versus the requirements for such set-ups, while section 4 discusses different sampling techniques in case of at-line sampling. In section 5, different techniques per monitoring approach, are evaluated based on the previously defined requirements. This section will go more in depth about specific measuring devices to monitor the key process variables. Section 6 and 7 discuss models and soft sensors, which can be used for optimization and control of a fermentation process. Finally, the discussion evaluates the applicability to the case study and discusses an optimal strategy for the monitoring of cellulose to ethanol fermentation.

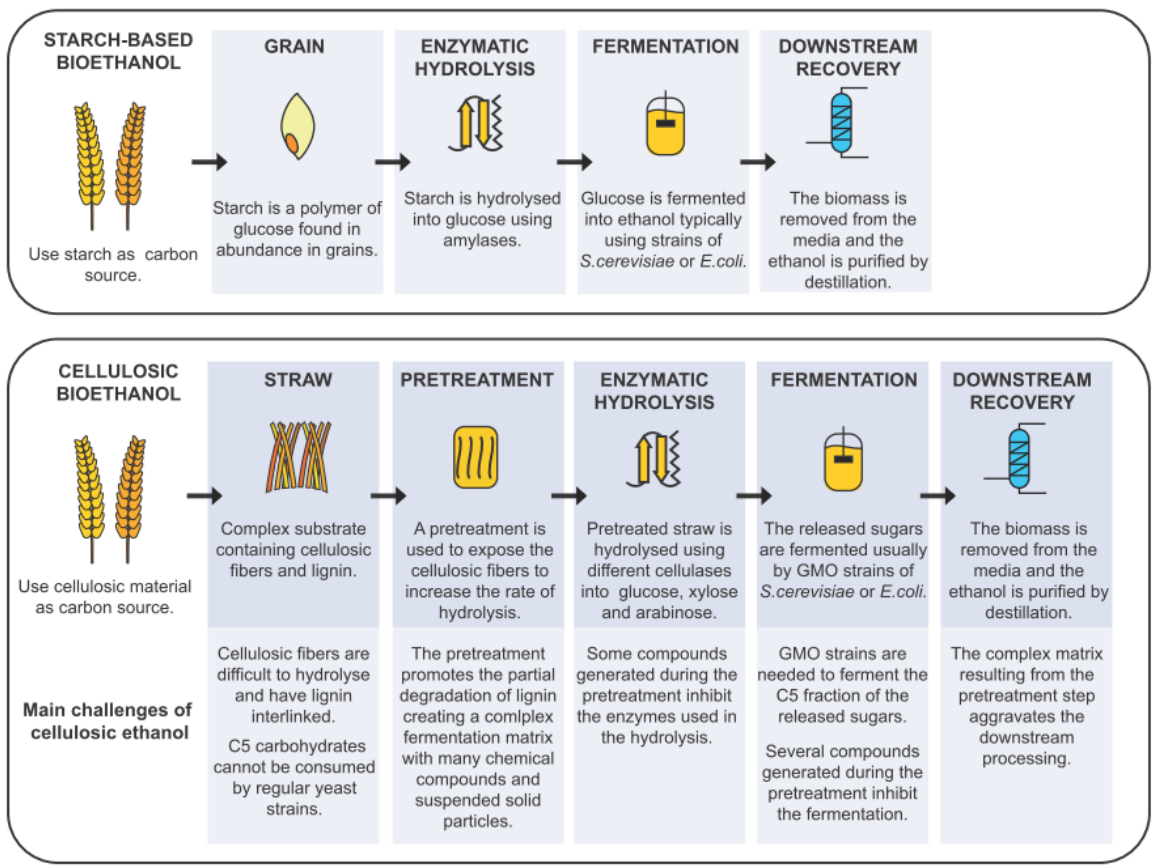

**Figure 1: Differences between non-cellulosic and cellulosic ethanol production. Note that both, non-cellulosic and cellulosic ethanol can also be produced from other sources such as sugar beets or wood chips respectively.**

## 2   The Cellulosic Ethanol Fermentation Process

The process to produce cellulosic ethanol typically consists of four consecutive steps: the pretreatment of the lignocellulosic material, the enzymatic hydrolysis of the pretreated material, the fermentation of the hydrolysate and the separation processes (**FIGURE 1**) [13]. In the pretreatment, the lignocellulosic fibers are broken down to smaller pieces, and exposed to increase the hydrolysis rate in the following step. Several methods are available for the pretreatment, most of them including high temperatures and pressures, and pH variations by addition of acid or base. Some conventional pretreatment methods are acid hydrolysis, steam explosion or ammonia treatment [13]. The choice of a specific pretreatment strategy will have an impact on the downstream processing, the hydrolysis and the fermentation steps, and may raise different challenges for the implementation of analytical methods to monitor the fermentation process, which must be considered. The enzymatic hydrolysis is the step in which the fibers of lignocellulose are enzymatically hydrolyzed to release the monosaccharides. In some cases, the enzymatic hydrolysis and the fermentation are performed simultaneously (simultaneous saccharification and fermentation, SSF) and in some other cases they are done consecutively (separate hydrolysis and fermentation, SHF) [13,14]. The performance of the hydrolysis will also have an impact on the fermentation, as it determines the concentration of fermentable sugars. The system considered in this case study is the fermentation step in a separate hydrolysis and fermentation process.

The fermentation for the production of cellulose-based ethanol usually consists of a batch phase, followed by a fed-batch phase, and finally another batch phase. A fed-batch operation typically starts and ends with a batch phase[15]. In the first batch phase, the cells grow at a maximum growth rate, and the cell density increases significantly. Cell and process characteristics define this growth rate. During the fed-batch phase, a feed stream enters the reactor increasing the volume in the reactor. In the reactor, anaerobic conversion of the substrates to product and biomass takes place with a rate dependent on the microorganisms used and the process characteristics (Figure 2). During the fed-batch phase, the conversion rate is limited by the feed rate and the detoxification of inhibitors. In effect, the admissible feed rate is generally limited by the presence of inhibitors in the feed and cannot proceed faster than the capacity of the micro-organism to detoxify the medium. It is therefore important that the amount of inhibitors is monitored during the fed-batch phase as a means to maximize the feed rate and productivity. During the final batch phase the consumption, production, and growth rates are not controlled. The capacity is defined by the host organism and the process characteristics, such as pH and temperature, and the presence of inhibitory compounds. The main cost components of this fermentation process are the feedstock, utilities, and capital cost [11]. It is therefore desirable to utilize as much of the feedstock as possible for ethanol production, so a high ethanol yield is required. Furthermore, to minimize the utilities cost, a high productivity is desired to minimize the fermentation time.

For this case study, it is assumed that the carbon source originates from wheat straw, which yields mainly glucose, xylose, furfural, 5-HMF, acetic acid, and lignin after pre-treatment and enzymatic hydrolysis [11]. In the current study, it is assumed that the yeast, which is a genetically engineered strain, can consume glucose and xylose simultaneously [16]. The productivity of the process is mainly dependent on the xylose consumption rate, as this is the rate-limiting step in mixed glucose/xylose fermentation. Furfural is a major inhibitor of yeast [17], and it is therefore important to keep this concentration low throughout the fermentation process. Acetic acid is also a major inhibitor, but the inhibition effect of this compound depends on the pH as only the unionized (neutral) form is inhibitory. This indicates that pH control is of importance for the process. As the fermentation is run without gas sparging, oxygen will be present in the beginning of the fermentation. This is important to monitor, as the presence of oxygen decreases the ethanol yield, as ethanol is produced under anaerobic conditions. It is therefore desired that the oxygen has been consumed before the fed-batch phase starts. The most important variables of cellulose to ethanol fermentation are therefore the carbon sources glucose and xylose, the product ethanol, the inhibitors furfural and acetic acid, carbon dioxide and oxygen, and the pH. These variables can be monitored in real-time by either direct measurement or indirect modeling techniques. The monitored variables can then be used in a model for optimization and control, as shown in Table 1 and Table 2, where respectively the process objectives and different risks and solutions associated with cellulosic ethanol fermentation are shown.

**Table 1: Monitoring targets to achieve process objectives**

| Monitoring target | Objective | | |
| --- | --- | --- | --- |
| | High yield | High productivity | Detect contamination |
| pH | - | - | Reactor |
| Temperature | - | - | - |
| Cell biomass | - | Reactor | Reactor |
| Glucose | Reactor/inlet | - | Reactor/inlet |
| Xylose | Reactor/inlet | - | - |
| Ethanol | Reactor/off-gas | - | - |
| Acetic acid | Reactor/inlet | Reactor/inlet | - |
| Furfural | - | Reactor/inlet | |
| 5-HMF | - | Reactor/inlet | |
| Lactic acid | - | - | Reactor |
| $CO_2$ | Off-gas | Off-gas | - |
| $O_2$ | Reactor/off-gas | - | - |

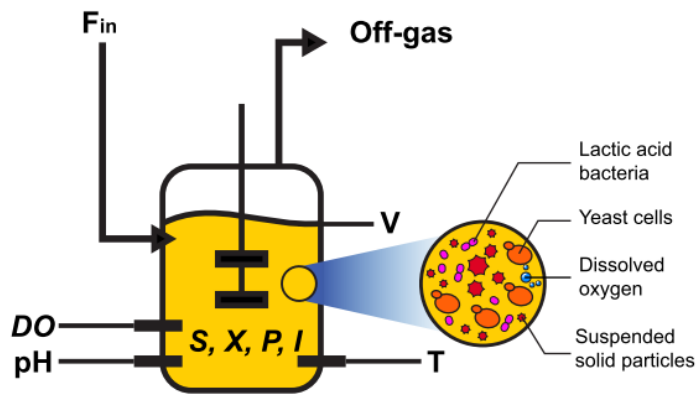

**Figure 2: Schematic overview of a fed-batch reactor with a feed rate $F_{in}$. The components in italics indicate the uncertainties in the process. These are the substrate (S), product (P), biomass (X), dissolved oxygen (DO), and the possible presence of a contamination. The feed rate $F_{in}$ is known and controlled. The off-gas composition is not known, but as it is an indirect indication of the state of the process, it is not a direct uncertainty in how the fermentation behaves. The volume (V), pH, and temperature (T) are usually monitored and controlled, and therefore not uncertain.**

**Table 2: Overview of risks associated with cellulose to ethanol fermentation.**

| Risk | Consequence | Monitoring targets to diagnose and detect | Possible solution |
|---|---|---|---|
| **Contamination by lactic acid bacteria (LAB)** | The growth of LAB depletes some of the glucose and xylose decreasing the ethanol yield | pH, biomass or component dissolved in the liquid (fast detection desired) | Early stop of the batch in order to not use all the substrate |
| **Inhibition by furfural and acetic acid** | The ethanol production rate decreases significantly until detoxification | Components dissolved in the inlet (with constant inlet, slow detection sufficient) and dissolved in the reactor (fast detection required) | Control of the feed rate, to ensure the concentrations of inhibitors remain under a certain threshold |
| **Presence of oxygen** | The ethanol yield will decrease, as ethanol is produced in anaerobic conditions | The off-gas, or detection in the liquid (fast detection required) | The oxygen will need to be consumed |

## 3   Key process variables

In this section, the added value of the monitoring of each key process variable will be evaluated in terms of what a monitoring strategy of different variables can add to the total quantity of process data that can be analyzed. Figure 2 gives an overview of the uncertain elements in cellulose to ethanol fermentation, which are shown in italics. A comparative table of the evaluation results can be found in Table 3, where each monitoring step has been assigned a number of points, depending on how much the measurements contribute to the analysis of the key variables. While Table 1 describes why components are measured, Table 3 describes how they are measured. Temperature and pH, which are standard monitoring techniques, are set at zero points. The other techniques are pointwise compared to the added value of temperature and the pH. The next few paragraphs will focus on how the table and figure are linked, and how the system is graded. The end rankings, which were reviewed by an industrial panel in ØRSTED (Denmark), with plenty of experience in operating a cellulosic ethanol demonstration plant, are a result of combining an extensive literature study, including academic research and published patents, and the authors' experience with monitoring and control. The targets addressed are monitoring the off-gas, the components dissolved in the feed stream, the components dissolved in the reactor, the biomass concentration, and detecting contaminations, such as the occurrence of lactic acid bacteria.

## 3.1 Temperature and pH

The most basic approaches to monitor a fermentation process are through the temperature and the pH. Most fermentation processes are run at constant pH with a relatively loose control. As carbon dioxide is produced along the fermentation, base is added to keep the pH constant. Under normal circumstances, the addition of base to the reactor at a relatively constant pace would indicate stable growth and ethanol formation. However, an abnormally large addition of base is an indication of a contamination with undesired lactic acid bacteria, as the production of lactic acid substantially acidifies the medium [18]. When a contamination is detected, the most convenient solution is to stop the process, as the substrate represents a major share of the production costs, and a contamination will take valuable carbon source away from ethanol production.

## 3.2 The off-gas

Measurements of the off-gas give the highest added value as a stand-alone method. It is possible to detect carbon dioxide, oxygen, and ethanol directly in the off-gas, and thus predict the concentrations in the liquid phase. This is usually done by using Henry's law, which is dependent on the process conditions, in particular temperature and, for carbon dioxide, pH. One can also indirectly monitor the growth rate, the total sugar consumption, and detect contaminations through mass balances and growth kinetics [19]. The ethanol concentration can give information on the process yield, while the process rates indicate the productivity of the process. Furthermore, monitoring the oxygen in the off-gas is important, as the presence of oxygen is unwanted in cellulose to ethanol fermentation.

### 3.3 The off-gas and components dissolved in the inlet

Combining off-gas measurements with measurements of the components dissolved in the inlet can give, additionally, the xylose and glucose concentrations in the liquid, as these can be estimated through mass balances and growth kinetics [20]. The off-gas provides feedback information about the rate of consumption/growth whereas the inlet measurements give feedforward information about the actual substrate provided. This also allows better estimation of the biomass concentration, as compared to solely off-gas measurements. Another advantage is that the inhibitory components entering the reactor are directly monitored, which has the result that the feed rate can be manipulated, to maintain a low concentration of inhibitors in the reactor during the fed-batch phase.

### 3.4 The off-gas and components dissolved in the reactor

When combining off-gas measurements with measurements of the components dissolved in the reactor, no predictions are needed to acquire these data, and real-time information of the actual state of the process can be obtained. On the other hand, not having any measurements of the inlet is a disadvantage because characterizing the inlet is important in general control of fermentations, especially when the inlet can be a potential source of disturbances. In this strategy, such disturbances would only be measured inside the reactor. During the fed-batch phase, the only manipulated variables are the feeding rate, the addition of base or to stop the batch and start all over again. While the pH is often maintained within certain bounds (at the expense of using base, which is expensive), the feeding rate can be adjusted to keep the concentration of inhibitors inside the reactor below a threshold. In this regard, the difference between monitoring the components dissolved in the reactor or in the inlet would be that the former would allow to control the feeding rate based on actual measurements, while the latter would depend on the prediction of how fast the cell culture can detoxify the inhibitors. Also, by monitoring compounds dissolved in the reactor it would be possible to directly measure the concentration of lactic acid, which would allow to early detect contaminations by lactic acid bacteria and to stop the batch on time.

## 3.5 The off-gas and the biomass concentrations

Another option is combining the off-gas measurements with the monitoring of the biomass concentration. This will not yield direct concentrations of glucose, xylose, ethanol, and furfural, but with the right measuring method contaminations could be observed directly. This is the only beneficial aspect of monitoring the biomass concentration instead of the before mentioned monitoring schemes, although as will be described in section 5.3, so far no applications are available that can distinguish cells on-line in industrial scale. The effect of inhibitory compounds can be seen in the biomass activity, but there is no knowledge of the amount of inhibitors that are present. This makes control of especially the feed rate significantly more complex.

## 3.6 Components dissolved in the inlet and in the reactor

If off-gas measurements are not possible, one could also measure the components in the inlet and in the reactor. This does not change the added value compared to the previous two mentioned methods, but measuring components dissolved in the liquid phase is more complex than off-gas measurements. Section 5 will elaborate in more detail on these differences.

## 3.7 Addition of multiple monitoring methods

Increasing the number of monitoring methods to three or four increases the added value, as different measurements will add more direct data. However, it should be noted how much additional monitoring approaches contribute to the total amount of information obtained from combining hardware and software sensors, as soft sensors are often capable of analyzing what is going on in the reactor from less complex measuring methods, such as the off-gas composition measurement. Hardware sensors should be better capable of giving accurate information on the current reactor state. However, this is only true if the sensors can measure all components of interest, are accurate and not subjected to interference. Furthermore, fast response times are beneficial for fast control, but the techniques should not be too expensive. Biomass concentration measurement techniques should be able to detect contaminations and distinguish between viable and non-viable cells to be of any extra value. Most techniques will need soft sensors for calibration and to convert the measured data into valuable information. The complexity of the calibration, maintenance and data analysis differs per technique, and this can be of importance when considering that factories are often built in remote areas, where expert knowledge will not always be available at all times. These considerations will be taken into account in section 5, where equipment is discussed.

Table 3: The added value of (combinations of) different monitoring strategies of key process variables. -: does not monitor, +: monitors indirectly, ++: monitors indirectly through different models, +++: monitors directly. Each plus counts as one point, while the points of the standard setup (pH and temperature measurements) are subtracted from the total amount of gained points for each monitoring strategy.

|  | Monitor components in the off-gas | Monitor components in the reactor | Monitor components in the inlet | Monitor the biomass concentration | Detect contaminations | Total score |
|---|---|---|---|---|---|---|
| pH + Temperature | - | - | - | - | + | 0 |
| Off-gas | +++ | + | - | + | + | 5 |
| Off-gas + inlet | +++ | ++ | +++ | ++ | + | 10 |
| Off-gas + reactor | +++ | +++ | - | ++ | ++ | 9 |
| Off-gas + biomass | +++ | + | - | +++ | +++ | 9 |
| Inlet + reactor | - | +++ | +++ | ++ | ++ | 9 |
| Off-gas + inlet + reactor | +++ | +++ | +++ | ++ | ++ | 12 |
| Off-gas + inlet + reactor + biomass | +++ | +++ | +++ | +++ | ++ | 13 |

## 4 Sampling

Real-time measurements can be performed either in-line, on-line or at-line (Figure 3) [21]. With in-line monitoring, the measurements are performed directly inside the reactor without removing or diverting the sample from the process stream. On-line and at-line measurements, in contrast, take place outside the reactor. While the sample is diverted and may be returned to the reactor (e.g. analysis through a flow cell) for on-line measurements, the sample is removed when performing at-line measurements. In order to maintain real-time measurements, on/at-line methods need to be automated for industrial applications. There is, therefore, a need for a reliable sampling technique, connected to one or multiple pre-treatment devices, and subsequently the measuring device. The pre-treatment devices often include filtration units to remove the suspended solid particles and flow systems to prepare the samples (e.g., to dilute or stain them). A promising automated pre-treatment method is cross-flow filtration, where a constant flow through a hollow fiber keeps solid particles from clogging the membrane [22,23]. This method has been used by Meschke *et al.* [22] in combination with high-performance liquid chromatography (HPLC), and by Rocha and Ferreira [23] with an amperometric biosensor. Also, the wastewater treatment sector applies cross-flow filtration in order to remove particles from the water or retain biomass in the reactor [24,25]. Another type of automated sampling techniques which is being developed is applied in the BioScope [26]. The BioScope can be used for experimental research of microbial kinetics in a fermentation, in which rapid sampling is desired. However, so far, this technique is developed for experimental research, and not for industrial applications. Automated sampling devices combined with a sample preparation system have also been described for the application of flow cytometry [27]. A general challenge for an automated sampling system is that sterility in the reactor needs to be maintained. However, for the case of cellulosic ethanol production, this is not an issue as the reactor is operated under non-sterile conditions.

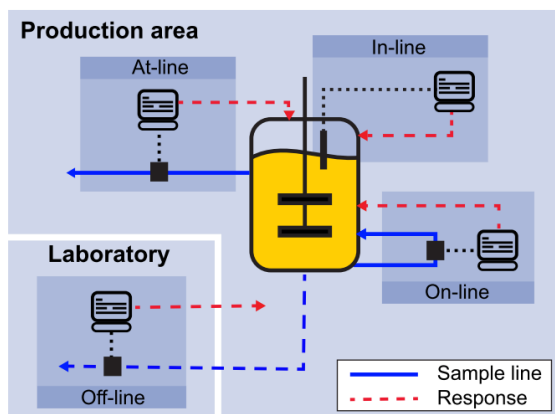

Figure 3. Conceptual approaches to real-time monitoring according to the guidance for industry PAT — A Framework for Innovative Pharmaceutical Development, Manufacturing, and Quality Assurance [21]. PAT: process analytical technology.

# 5 Sensors

This section will evaluate different measuring techniques for the monitoring approaches discussed in section 3.

## 5.1 The off-gas analyzer

Previously, an extensive review has been written on different methods to measure the off-gas composition, which is considered as a continuous measurement [19]. For this study it was chosen to only focus on techniques that can measure all gas components of interest, as combining different gas monitoring methods in tandem will be more expensive [3]. The only two techniques that can measure all components of interest, carbon dioxide, oxygen, and ethanol, are electronic noses and mass spectrometry, as these methods are capable of measuring a broad spectrum of volatile components in the off-gas. The electronic nose works as a semiconductor, where the resistance of sensors changes when exposed to volatile organic compounds (VOC) or gases. An electronic nose consists of multiple sensors with high sensitivity, but a slow response time. Furthermore, pattern recognition algorithms are needed to analyze the obtained data. Another issue is that background gases such as water vapor can interfere with the measurements [28]. A solution for this has been proposed, where samples were dehydrated before injection into the electronic nose [29]. A significant amount of research is being conducted on the electronic nose, but the main applications are in food technology, and most of the applications are still performed on lab scale. Mass spectrometry (MS) on the other hand is a well-established method [30] capable to quantify a broad range of substances with high accuracy, typically from 100% to a few parts per million. One can choose for quadrupole MS, which is the cheaper option, or magnetic MS, which is more expensive, but also more stable and offers a higher resolution.

## 5.2 Components dissolved in the liquid

In Table 5, the different techniques to monitor components dissolved in the liquid for both the inlet and the reactor are compared. The techniques evaluated are in-line, on/at-line near-infrared spectroscopy (NIR), mid-infrared spectroscopy (MIR), Raman spectroscopy, UV-Vis spectroscopy, biosensors, and HPLC. Fluorescence spectroscopy is not considered in this section because the key components dissolved in the liquid (i.e. glucose, xylose, ethanol, acetic acid, lactic acid, furfural and HMF) are not fluorescent. The evaluation is based on the following eight requirements of measured components: sensitivity, accuracy, drift, calibration and data analysis, sample preparation, response time, industrial availability, and costs. In order to compare the potential of each technique, a scoring matrix is introduced which is made considering each of the previous criteria. The scoring matrix aims at reflecting the applicability and complexity of each method to provide a better understanding of the possibilities of each technique. The requirements were based on previous literature [12] and discussions with industry. An example of how the scoring matrix is done is provided for the first criterion (measured components) in Table 4. A detailed explanation of the development of the scoring matrix for the remaining seven criteria is provided in the supplementary material.

**Table 4. Scoring matrix to evaluate the capabilities of the different methods to monitor the key compounds of the cellulose to ethanol fermentation. A method capable of monitoring all the relevant compounds would receive a score of 3, while a method unable to monitor any of the compounds would receive a score of 0.**

|  | On-line NIR | In-line NIR | On-line MIR | In-line MIR | On-line Raman | In-line Raman | On-line UV-Vis | Biosensors* | At-line HPLC |
|---|---|---|---|---|---|---|---|---|---|
| Glucose | Yes | Yes | Yes | Yes | Yes | Yes | No | Yes | Yes |
| Xylose | Yes | Yes | Yes | Yes | Yes | Yes | No | Yes | Yes |
| Ethanol | Yes | Yes | Yes | Yes | Yes | Yes | No | Yes | Yes |
| Acetic acid | Yes | Yes | Yes | Yes | Yes | Yes | Yes | Yes | Yes |
| Lactic acid | No | No | Yes | Yes | Yes | Yes | Yes | Yes | Yes |
| Furfural | No | No | No | No | No | No | Yes | No | Yes |
| HMF | No | No | No | No | No | No | Yes | No | Yes |
| **Total score** | **1** | **1** | **2** | **2** | **2** | **2** | **1** | **2** | **3** |

### 5.2.1 Vibrational spectroscopy

Vibrational spectroscopy (UV-Vis, NIR, MIR and Raman spectroscopy) is a group of analytical techniques that allow a fast detection of several compounds directly from the fermentation media without the need for sample preparation. The primary challenge for the application of vibrational spectroscopy to monitor cellulose to ethanol fermentation is the high content of suspended solid particles derived from lignin and biomass. These particles interfere with the light, reflecting and scattering it. This limits the implementation of vibrational transmission spectroscopy to on-line or at-line modes only, where a filtration unit is added before the spectroscopic analysis [31]. In contrast, reflectance vibrational spectroscopy (mainly attenuated total reflectance (ATR) and diffuse reflectance [32]), and backscattered Raman spectroscopy do not depend on the light transmitted through the media but on the light reflected or backscattered by the media, making these methods more suited for in-line monitoring cellulose to ethanol fermentations [33–35]. Despite the advantages of reflectance and backscattered spectroscopy, the interference between the particles and the light still entails extensive data pre-treatments and results in lower accuracy and sensitivity [35,36]. For this reason, vibrational spectroscopy methods performed better in the evaluation for on-line or at-line modes than for in-line modes.

Among the different vibrational spectroscopy techniques, near-infrared (NIR) spectroscopy is the most mature and well-established method [14,35,37], and it has been applied to monitor a wide variety of fermentations [33,38–40] including cellulose to ethanol processes [34,41,42]. Pinto *et al.* [41] used at-line transmission NIR to monitor the concentration of glucose and ethanol during cellulose to ethanol fermentation at lab-scale. Despite filtering the samples before analysis, the high interference of NIR with water and the highly overlapped spectra resulted in high prediction errors (6.60 g/L and 3.02 g/L for glucose and ethanol respectively). In another study, Sundvall *et al.* [42] used an on-line NIR probe (score of 11) in a demonstration-scale cellulose-to-ethanol plant (EPAB/SEKAB E-Technology, Sweden) to monitor the concentration of total sugars, glucose, ethanol, and suspended solids. Despite the good correlation between the off-line and the on-line samples, the reported concentration ranges were quite high (17-30 g/L and 2-40 g/L for glucose and ethanol respectively) and more sensitive measurements would be needed for accurate monitoring of the fermentation. Austin *et al.* [34] monitored the concentration of total sugars, glucose, and ethanol in a 23 m$^3$ reactor using an in-line diffuse reflection probe (score of 11). The measurements were noisy due to the high concentration of solid particles but in accordance with the off-line measured samples, giving valuable qualitative information about the process endpoint. In general, NIR spectroscopy has the advantage of being a robust method that can be implemented in, on or at-line, and requiring very little or no sample preparation. Although it is not as sensitive and accurate as other techniques, NIR delivers qualitative information that can increase the process knowledge. For these reasons, on-line and in-line NIR is given a score of 11.

Mid-infrared spectroscopy (MIR) offers a higher accuracy and a larger number of variables to be analyzed compared to NIR [5,34,43]. Several implementations of MIR in cellulose to ethanol bioprocesses are reported in the literature [31,34,44]. Juhl *et al.* [31,44] used an at-line transmission system to monitor the concentration of glucose, lactic acid, glycerol, acetic acid, and ethanol. The samples were filtered prior to analysis in order to avoid the interactions with the solid particles. The predictions with MIR had a significant lower root mean square error of prediction (RMSEP) when compared to the ones obtained with NIR in a similar set-up (e.g., the RMSEP for glucose was 0.12% for MIR and 0.26% for NIR). In another study, Austin *et al.* [34] used an in-line attenuated total reflectance MIR (ATR-MIR) probe (score of 13) to monitor the glucose, xylose, lactic acid, acetic acid and ethanol concentration in a 23 m$^3$ reactor. Their results were directly compared with in-line diffuse reflectance NIR and showed that ATR-MIR had a significantly higher accuracy than NIR [34], allowing a better understanding of the dynamics of the fermentation. On-line ATR-MIR (score of 12) has also been applied to the hydrolysis step of starch-based ethanol production and brewing processes, which present similar challenges as cellulose-based ethanol production regarding suspended solid particles [45,46]. ATR-MIR has a shallow penetration depth in the sample media, making it more robust in media with suspended particles than transmission MIR. The main disadvantages of ATR-MIR are the fouling on the surface of the ATR crystal [31] and the high costs associated with the optical fibers required to transmit the signal. ATR-MIR scores higher than NIR spectroscopy (12 and 13 for on-line and in-line respectively) due to the higher sensitivity and accuracy, and due to the potential to measure lactic acid, a crucial compound to detect contaminations.

Raman spectroscopy is an attractive method foremost because there is, unlike for NIR and MIR, no water interference. Additionally, Raman spectra are better resolved and require less modeling efforts than NIR and MIR [47,48]. However, the Raman signal is relatively weak and attenuated mainly by the suspended solid particles and by the background fluorescence emitted by lignin [48], altogether, limiting its potential for in-line monitoring. Ewanick *et al.* [49,50] used on-line Raman spectroscopy (score of 12) to measure the concentrations of glucose and ethanol in a lab-scale cellulose to ethanol fermentation (1.3 L). In order to avoid the interference with suspended solid particles, the fermentation medium was filtered prior to the fermentation. The concentration of glucose and ethanol were monitored with a prediction error of 1 g/L. Also at lab-scale, Iversen *et al.* monitored the concentration of glucose, ethanol and acetic acid using in-line Raman spectroscopy (score of 12) [47,48]. To account for the reduction of fluorescence caused by the suspended solid particles, Iversen *et al.* included an internal standard as a correction factor [51]. Despite the efforts to minimize the effect of the solid particles, their research showed that accuracy of Raman spectroscopy improves when lignin particles are removed before the measurement, which on full-scale could be achieved by using an automated sample port in combination with a filtration or sedimentation step. In spite of the potential of Raman spectroscopy as analytical technique, the expensive material and the lack of relevant industrial implementation lead to suggest a final score of 12 for both, on-line and in-line Raman spectroscopy.

UV-Vis spectroscopy is often not considered as a method for real-time monitoring of fermentations because it cannot detect many key compounds (e.g. glucose or ethanol) and because the light scattering caused by the suspended solid particles dominates the absorption process [35,52]. However, in the context of cellulose to ethanol fermentation, the technique gains special relevance because many of the inhibitors present in lignocellulosic hydrolysate including furfural, HMF or acetic acid absorb in this region [53]. Pinto *et al.* [53] used at-line UV-Vis spectroscopy to quantify the concentration of furfural and HMF from filtered samples, attaining a high sensitivity and low prediction errors (RMSEP of 0.375 g/L and 0.041 g/L for furfural and HMF respectively). UV-Vis is a useful method to quickly detect inhibitory compounds and lactic acid (useful to detect contaminations), in an inexpensive manner. For this reason, UV-Vis gets an overall score of 10.

### 5.2.2  Biosensors

Biosensors (total score of 10) in general use enzymatic reactions to monitor concentrations of specific components [54]. The way the reactions are monitored differs per type of biosensor. The most widely known biosensor is the amperometric glucose sensor, which is used by diabetes patients to measure glucose levels in the blood [55]. In the biosensor glucose oxidase converts glucose to hydrogen peroxide ($H_2O_2$), which reacts with specific compounds in the sensor and generates, in the case of an amperometric biosensor, a current, which is measured. Ethanol can be monitored by the same principle through the use of alcohol dehydrogenase [56]. The measurement of xylose can be monitored simultaneously with glucose by the YSI 2700 SELECT probe (YSI Life Sciences, Yellow Springs, Ohio, USA), but sample filtration and dilution are required. Concentration ranges of 0.05 g/L – 9 g/L and 0.5 g/L – 30 g/L were reported for glucose and xylose, respectively (YSI Life Sciences, 2008). Amperometric sensors for the detection of lactic acid have been developed and applied to monitor malolactic fermentations [58]. This can be used to detect lactic acid bacteria. The measurements are fast, sensitive, and have a high selectivity. However, the sensors have limited long term stability and drift is encountered [3]. This happens in the time range of days to months, depending on the sensor [56]. Electrochemical sensors to monitor the concentration of acetic acid in fermentations have also been described in the literature [59]. There are also no reports on the measurement of furfural through biosensors, but as the sensors work enzymatically, this should theoretically be possible. The technique has not yet been applied on industrial scale, which forms an indication that there is still considerable development work needed.

### 5.2.3 High-performance liquid chromatography (HPLC)

The most widely used and known method of measuring specific components is HPLC, which is commonly used as reference measurement to calibrate other monitoring methods. For at-line applications, a flow injection system that withdraws, filters and prepares the sample is required so that only particle-free liquid is analyzed by the HPLC [2]. This adds complexity to the set-up and increases its costs and operational time. Furthermore, the HPLC columns need to be washed regularly to guarantee that one obtains reliable results. In order to reduce the complexity of the set-up, it is desired to use a single chromatographic column able to analyze as many relevant compounds as possible. In the context of cellulosic ethanol production, the simultaneous quantification of sugars (glucose and xylose), ethanol, acetic acid and common inhibitors (HMF and furfural) is challenging and slow due to their different chemical properties and concentration ranges [60]. The simultaneous quantification of the previously mentioned compounds has only been reported using an Aminex HPX-87H column and requires between 40 to 55 minutes for one analysis depending on the mobile phase [60–62]. Faster analysis (up to 15 minutes) would be achieved using different columns, but it would increase the costs of the set-up and the complexity of the operation [5,22,61,63]. At-line HPLC gets a high score as an analytical tool (as it can measure all relevant compounds with high sensitivity and accuracy, and a small drift), but it is somewhat challenging to automate, requires sample preparation and has a slow response time (total score of 12).

Table 5: Overview of all the techniques discussed to monitor components in the liquid phase. Scores from 0 to +++ are given for each criterion, 0 indicating a negative effect and +++ indicating a positive effect. The costs are evaluated with scores from --- to 0, --- indicating more costly and 0 less costly.  A thorough description of the scoring system is provided in the supplementary material.

| | Measured compounds | Sensitivity | Accuracy | Drift | Calibration & data analysis | Sample preparation | Response time | Industrial implementation | Costs | Total score |
|---|---|---|---|---|---|---|---|---|---|---|
| **On-line NIR** | + | + | ++ | + | + | ++ | ++ | ++ | - | 11 |
| **In-line NIR** | + | + | + | + | 0 | +++ | +++ | ++ | - | 11 |
| **On-line MIR** | ++ | ++ | ++ | + | + | ++ | ++ | ++ | - - | 12 |
| **In-line MIR** | ++ | ++ | + | + | 0 | +++ | +++ | ++ | - | 13 |
| **On-line Raman** | ++ | ++ | ++ | ++ | + | ++ | ++ | + | - - | 12 |
| **In-line Raman** | ++ | ++ | + | ++ | 0 | +++ | +++ | 0 | - | 12 |
| **On-line UV-Vis** | + | + | ++ | ++ | + | ++ | ++ | 0 | - | 10 |
| **Biosensors** | ++ | ++ | ++ | 0 | ++ | + | + | 0 | 0 | 10 |
| **At-line HPLC** | +++ | +++ | +++ | ++ | ++ | 0 | 0 | + | - - | 12 |

## 5.3 The biomass

Monitoring biomass in cellulose to ethanol fermentation is a significant challenge foremost because the conventional methods used in other fermentation processes (e.g., optical density probes or infrared spectroscopy) fail at differentiating cellular biomass from the suspended solid particles and therefore are not suitable for lignocellulosic ethanol fermentations [6]. Moreover, standard methods to assess cell culture viability (e.g., methylene blue test) cannot be applied due to the dark color of the media [64]. In biomass monitoring, unlike in methods to monitor compounds in the liquid, the samples cannot be filtered prior to  analysis because that would also remove the cells. In this section, different methods to monitor the biomass concentration are discussed and evaluated regarding their ability to differentiate between biomass and solid particles, to assess the cell culture viability, to detect contaminations, sample preparation, calibration, and data analysis, industrial availability and costs. An overview of the evaluation can be found in Table 6, and a detailed explanation of the scoring system is provided in the supplementary material.

### 5.3.1 Multi-wavelength fluorescence spectroscopy

Fluorescence spectroscopy (total score of 7) can monitor biological compounds such as NADH, tryptophan, and riboflavin [6]. These compounds are closely related to the generation of cells and can be used as indirect measurements of biomass [65–68]. Multi-wavelength fluorescence spectroscopy produces three-dimensional data sets (time, excitation spectra and emission spectra) which are analyzed using advanced chemometric methods (typically using parallel factor analysis, PARAFAC [65–67]). By using these models, it is possible to resolve the pure spectra of each fluorophore from the mixture, making multi-wavelength fluorescence more robust to changes in the composition of the media and to the background fluorescence emitted by lignin [65,66,69]. In addition, similarly to other spectroscopic techniques, fluorescence spectroscopy is also affected by the high content of suspended solid particles. Multi-wavelength fluorescence has previously been used to monitor ethanol fermentations at lab-scale, but there are no reports of utilizing fluorescence spectroscopy for cellulose-based ethanol production. The BioView fluorescence spectrometer (Delta, Hørsholm, Denmark) claims to be applicable in industrial settings [70] but has to our knowledge not been used for the monitoring of ethanol production from lignocellulosic biomass at pilot or even larger scale.

### 5.3.2 Biocalorimetry

A biocalorimeter (score of 8) monitors biomass growth based on the metabolic heat, which is calculated from all the heat flows concerning the reactor [71]. The main advantage of this technique is that the equipment needed, mainly temperature probes and flow meters, is cheap [72]. A direct relation was even found between the consumption of cooling water and the metabolic heat generation in an industrial-sized bioreactor of 100 m$^3$, where the biomass concentration could be estimated more accurately using the cooling water consumption data than from elemental and electron balances [72]. In fact, as the scale of the reactor increases, smaller influences such as heat loss to the environment and noise become less significant. This method monitors the biomass concentration indirectly through heat balances, in a similar way as it can be monitored through a carbon balance, although no distinction between cell types can be made. The initial biomass concentration needs to be known to estimate the concentration over time from the metabolic activity. Response times are between 1 and 2 minutes [5].

### 5.3.3 Flow cytometry

Flow cytometry (score of 8) is an at-line method to characterize and count cells through light scattering and fluorescence [73]. It can monitor the biomass concentration accurately and allows to distinguish between viable cells, non-viable cells, and other types of biomass [74]. Flow cytometry is expensive, but it has been applied on large scale and many different devices are available [73]. In addition, several approaches have been developed to automate the sampling procedure, dilution, and staining of the cells via flow injection systems, thereby reducing the required labor and allowing the design of control strategies based on the physiological properties of the cell culture [75–79]. The main drawback of flow cytometry in cellulose to ethanol fermentations are the suspended solid particles, which cannot be filtered and can only be differentiated from the biomass via expensive fluorescent stains and not via light scattering. Apart from that, accuracies have been reported to be good enough up to a concentration of $2 \cdot 10^6$ cells/mL, which means that the samples will need to be diluted. As dilutions also increase the measurement error, it was observed that flow cytometry can only work well with a total concentration of up to $30 \cdot 10^6$ cells/mL [75]. The dilution steps will also increase the time needed for sample preparation. Sampling results can be obtained every 15 minutes [3].

### 5.3.4 Dielectric spectroscopy

Dielectric spectroscopy (score of 12), the most advantageous technique according to this study (Table 6), can monitor viable cells in-line by using an electric field at different frequencies to characterize the capacitance and conductivity of the system. The applied electric field induces the polarization of viable cells only [80,81], and this is reflected in the capacitance of the system [82]. Since polarization is only induced in viable cells, this method has no interference with gas bubbles and dead cells [83]. Dielectric spectroscopy has been applied to monitor cell viability in different fermentations with concentration ranges reported to be between 0 g/L and 200 g/L [4,5]. Furthermore, this technique has also been applied to control fermentations based on the specific growth rate [84]. Bryant *et al.,* [80] applied dielectric spectroscopy to monitor the hydrolysis of pretreated lignocellulose in a simultaneous saccharification and fermentation (SSF) process. Wang *et al.* [64] combined dielectric spectroscopy with multivariate analysis to measure the viability of yeast during a fed-batch SSF. Despite the positive results, the method requires extensive calibration to account for the different process parameters that affect dielectric spectroscopy (e.g., suspended solids, ethanol concentration or conductivity of the media). Another advantage of this technique is that it is available for industrial use, as industrial brewing processes already apply dielectric spectroscopy [5].

### 5.3.5 Microscopy and image analysis

Microscopy combined with image analysis (score of 11) is an automatic cell counting method based on the identification of individual cells from pictures taken with microscopy from fermentation samples [85]. It was developed 30 years ago in the brewing industry, and it has significantly developed with the recent advances in machine learning and improvements in detection sensors (i.e., charge coupled devices) [85,86]. Image analysis has also been used to correlate several features (e.g., cell size or cell volume) with cell viability. Donnelly *et al.* [87] developed a method to predict the viability of cell cultures with the cell volume distribution and used it to calculate the pitch size in industrial fermentations. Belini *et al.* [88] used in-line microscopy combined with image analysis to monitor yeast growth in a lab-scale molasses-to-ethanol fermentation. By using classification algorithms, they were able to differentiate between yeast cells and other solid compounds present in the fermentation media (e.g., plant fibers, sugar crystals or gas bubbles). If the resolution of the microscope is high enough, this method can also be used to detect microbial contaminations, as suggested by Belini *et al.* [86].

Table 6: Overview of all the techniques discussed to monitor the biomass concentration. Scores from 0 to +++ are given for each criterion, 0 indicating a negative effect and +++ indicating a positive effect. The costs are evaluated with scores from --- to 0, --- indicating more costly and 0 less costly.  A thorough description of the scoring system is provided in the supplementary material.

| | Differentiate cells and particles | Assess cell viability | Detect contaminations | Sample preparation | Calibration & data analysis | Industrial availability | Costs | Total score |
|---|---|---|---|---|---|---|---|---|
| ***OD probes*** | 0 | 0 | 0 | ++ | +++ | ++ | 0 | 7 |
| **IR spectroscopy** | 0 | 0 | 0 | +++ | + | + | - | 4 |
| **2D fluorescence** | +++ | 0 | 0 | +++ | + | + | - | 7 |
| **Bio-calorimetry** | +++ | 0 | 0 | +++ | +++ | 0 | - | 8 |
| **Flow cytometry** | +++ | +++ | +++ | 0 | ++ | 0 | - - - | 8 |
| **Dielectric spectroscopy** | +++ | +++ | 0 | +++ | + | ++ | 0 | 12 |
| **Microscopy and imaging analysis** | +++ | +++ | +++ | ++ | 0 | + | - | 11 |

## 6 Previous modelling efforts

The previous sections evaluated what measurements add to the extent of knowledge of cellulose to ethanol processes, and what measurement equipment is actually available for full-scale bioreactors. Models will be needed to predict the yield and productivity from the available data. The use of models is beneficial to control the process and optimize at specific points, such as the feed rate. Furthermore, it is important to model the variables that are considered as risks in Table 2, namely if there is contamination, inhibition, or presence of oxygen. These risks can be monitored directly through measurements, as described previously or indirectly through modelling. This section will look into the available models that take into account the measurements that were previously shown to be important to monitor the yield, productivity, and risks. A list of the models that have been evaluated can be found in Table 7. The models evaluated in this study are all unstructured models with simplified kinetic expressions (containing only substrate, product, and biomass), as structured models, containing synthesis rates of enzyme and intracellular metabolite production are considered too complex for routine daily use in a production environment. An interesting observation is that only one of the models described takes carbon dioxide in the form of total inorganic carbon into account [89], while the monitoring of this compound in the off-gas can relate significantly to the process characteristics. However, as cellulose to ethanol fermentation is not aerated or sparged, it is relatively difficult to monitor the gas flow rate out of the reactor and relate it to the dissolved $CO_2$ concentration. Therefore, it would be necessary to compare it with previous fermentations, and generate a relation based on experience. All evaluated models contain inhibition functions, often with Monod type kinetics. All studied models take product inhibition into account. Substrate inhibition and furfural inhibition, which was previously mentioned to be a strong inhibitor (see section 1), are also often modelled. In fact, Navarro *et al.* [90] only used furfural as state variable to describe the process. Monitoring the inhibitory compounds is important in a cellulose to ethanol fermentation, as the amount of inhibitory compounds in the reactor can be controlled through the feed rate. With the exception of the model published by Navarro *et al.* [90], all models contain at least the substrate and product as state variables, while the cell biomass is often present. These state variables are important to model the yield and productivity of the fermentation. Furthermore, sudden changes in yield and productivity can indicate the presence of inhibitory compounds or a contamination. In the case of Hanly and Henson [91], Palmqvist *et al.* [92], and Mauricio-Iglesias *et al.* [89], other major components present in the reactor are also included. In general, the more components are added in a model, the more accurate balance equations can be applied, and the more time will be spent on model development as well. Balance equations use relationships that are derived from theory or experiments to estimate states from measurements [93].

**Table 7: Overview of the process models researched in this study.**

| Reference | Model type | State variables | Substrate | Product | Inhibitors |
|---|---|---|---|---|---|
| **Krishnan *et al.* [16]** | Monod form expression for substrate inhibition. Two constant model for product inhibition | Glucose<br>Xylose<br>Ethanol<br>Cell biomass | Glucose<br>Xylose | Ethanol | Glucose<br>Xylose<br>Ethanol |
| **Navarro [90]** | Polynomial inhibition | Furfural | Glucose | Ethanol | Furfural |
| **Palmqvist *et al.* [92]** | Mechanistic model taking stoichiometric balances and metabolic pathways into account | Glucose<br>Ethanol<br>Cell biomass<br>Glycerol<br>Furfural | Glucose<br>Furfural | Ethanol<br>Glycerol<br>Furfuryl alcohol | Furfural<br>Glucose |
| **Zhang *et al.* [94]** | Monod type kinetics with competitive substrate inhibition for glucose and xylose, and an additional term for ethanol inhibition | Glucose<br>Xylose<br>Ethanol<br>Cell biomass | Glucose<br>Xylose | Ethanol | Glucose<br>Xylose<br>Ethanol |
| **Luong [95]** | Monod type kinetics with Levenspiel product inhibition | Ethanol (main component)<br>Glucose | Glucose | Ethanol | Ethanol |
| **Starzak *et al.* [96]** | Unstructured models for kinetics of cellular metabolism. Monod type kinetics for biomass growth and exponential ethanol inhibition. | Sucrose<br>Ethanol<br>Cell biomass | Sucrose | Ethanol | Ethanol |
| **Hanly and Henson [55]** | Monod type kinetics including several inhibition functions | Glucose<br>Xylose<br>Ethanol<br>Oxygen<br>Furfural<br>HMF<br>Furfuryl alcohol<br>2,5-bis-hydroxymethylfuran<br>Acetate<br>Cell biomass | Glucose<br>Xylose | Ethanol<br>Furfuryl alcohol<br>2,5-bis-hydroxymethylfuran<br>Acetate | Ethanol<br>Furfural<br>HMF<br>Furfuryl alcohol<br>2,5-bis-hydroxymethylfuran<br>Acetate |
| **Phisalaphong *et al.* [97]** | Monod type kinetics including | Sugar (cane molasses)<br>Ethanol | Sugar | Ethanol | Sugar<br>Ethanol |
| | several inhibition functions | Cell biomass | | | |
| **Pinelli *et al.* [98]** | Monod type kinetics with product inhibtion | Glucose Lactic acid Cell biomass | Glucose | Lactic acid | Lactic acid |
| **Athmanathan *et al.* [99]** | Monod type kinetics with the Levenspiel product inhibition function | Glucose Xylose Ethanol | Glucose Xylose | Ethanol | Ethanol |
| **Leksawasdi *et al.* [20]** | Monod type kinetics with substrate limitation and inhibition, and product inhibition | Glucose Xylose Ethanol | Glucose Xylose | Ethanol | Glucose Xylose Ethanol |
| **Mauricio-Iglesias *et al.* [89]** | Monod type kinetics including several inhibition functions | Reactor holdup Glucose Xylose Furfural Acetate HMF Ethanol Furfuryl alcohol Base conjugated cations Total inorganic carbon Cell biomass | Glucose Xylose | Ethanol | Glucose Xylose Ethanol Furfural Acetic acid HMF |
| **Wang *et al.* [100]** | Segregated model with 2 populations (active and inactive) controlled by acetic acid. Active type is a Monod type kinetics with glucose and xylose as substrate. Inactive type only consumes glucose. | Glucose Xylose Ethanol Acetic acid Cell biomass | Glucose Xylose | Ethanol | Glucose Acetic acid |

## 7    Soft sensors

Soft sensors are important for data analysis, process control, and process optimization. Data driven soft sensors are used to calibrate and interpret the data from measuring devices (hardware sensors), and to perform fault detection, from which deviating activity in the system can be found [101]. The most used soft sensors for this purpose are based on principal component analysis (PCA) decomposition and partial least squares (PLS) regression [35–37,67,101,102], which are applicable to linear relationships. For non-linear relationships, artificial neural networks (ANN) are often used. A challenge of ANN's is that they tend to get stuck in local minima [101]. For this reason ANN's need a significant amount of calibration data and tuning [103]. Soft sensors based on chemometrics, PCA for exploratory analysis and PLS regression are a mature technology and currently the most frequently applied tools in industry for monitoring fermentation processes. Furthermore, these soft sensors comply with the process analytical technology (PAT) initiative by the American Food and Drug Administration. These methods are very efficient for quality surveillance in order to detect if a particular process is following the intended production recipe. Hence, these tools provide insight into the current behavior of the principal components, but do not provide information which can be directly coupled with a first principles process model in order to predict or optimize future behavior.

Model-driven soft sensors on the other hand are applied to estimate variables from other monitored variables, to work as a backup for when hardware sensors fail, and to perform fault detection. The model-based soft sensors rely on first principles process models (balance equations for mass and energy as well as constitutive equations for e.g. reactions and transport) and on an algorithm that reconciles the available measurements with predictions by the model. This is also known as a filter or a state observer. Examples of such algorithms are Luenberger or Kalman filters or asymptotic observers [104,105].

Soft sensor technology has been utilized in the bulk chemical industry for decades but industrial applications in the biochemical industry are recent and under development [106,107]. The reasons for later utilization in e.g. fermentations can be several, among others, process-model mismatch, nonlinear dynamics, noisy measurements and that the development of state estimators of sufficient quality is troublesome for many industrial fermentation processes. Much of the research in state estimation focuses on ensuring the long-term (asymptotic) convergence of the developed algorithms. However, as the biochemical industry is dominated by batch and fed-batch processes (time limited), the ability of many popular state estimators to monitor bioprocesses is somewhat limited [104]. Furthermore, the instrumentation can be insufficient in order to have enough information available for the estimation. In industrial fermentation applications, spectroscopic methods dominate to a high degree, and these are not as straightforward to couple to the estimation scheme as direct measurements of e.g. temperature, pressure or pH, as is the case in classic chemical processes.

According to Luttmann *et al.* [108] soft sensors are mainly applied to determine the rate of oxygen consumption and carbon dioxide production, as well as the relationship between the two, the respiratory quotient (RQ) [109], but the number of applications at industrial scale is low. Furthermore, the RQ is not applicable to cellulose to ethanol fermentation, as there is no oxygen consumption. Mauricio-Iglesias *et al.* [89] explored the use of the continuous-discrete extended Kalman filter to estimate biomass, furfural and acetic acid by measuring glucose, xylose, ethanol and pH. The *in silico* results were promising as the estimation was reasonably good even in conditions of simulated contamination by lactic acid bacteria. So, to our opinion this is certainly a route that could be exploited further, for example for more standardized comparison of sensors, monitoring and control strategies *in silico*. Here, inspiration can be found in the wastewater treatment field, where benchmarking efforts aiming at *in silico* comparison of control strategies have been ongoing for almost 20 years now [110].

## 8    Discussion

This paper aimed to identify key variables to monitor in cellulose to ethanol fermentation. As cellulosic ethanol cannot yet compete with non-cellulosic ethanol regarding process economy, it is important to reduce the costs, which are mainly associated with utilities, substrate, biomass, and capital costs. Hence, an increase in profit can be achieved by increasing the yield and productivity as well as by running the fermentation in non-sterile conditions. However, to reach these objectives and to maintain the highest possible yield and productivity, monitoring and control are needed.

The current real-time monitoring methods used in the non-cellulosic ethanol industry (as in many other low-value, high-volume processes) consist of secondary measurements such as pH, turbidity, $CO_2$ in the offgas or temperature [111]. Although these measurements provide valuable information about the process, they do not directly relate to the state of the system, making it challenging to establish advanced control strategies. Similar to fermentation processes for non-cellulosic ethanol production, cellulosic ethanol fermentations are subject to fluctuations in the substrate composition that change the dynamics of the fermentation. Therefore, these processes would benefit from more advanced monitoring methods that can generate data that can be used for adjusting the operation of the process. When compared with non-cellulosic ethanol production processes, cellulose-based ethanol production is a more complicated process involving more phenomena such as inhibition, or a mixed substrate. In consequence, the monitoring methods typically used for the production of non-cellulosic ethanol fail in cellulosic ethanol production processes at providing real-time information, which would otherwise be useful for implementing control strategies. Additionally, advanced monitoring methods are required to improve the performance of cellulosic ethanol fermentations.

Models are needed to control and optimize the process. For reliable and accurate models, measurements are necessary. In the reactor, fast response times are also desired, as the process characteristics will constantly change. As the response times needed differ per process, it would be of value to investigate the actual response times needed in different processes.

Automatic controllers will also need real-time measurements as input. However, real-time monitoring of cellulosic ethanol fermentation is complex and troublesome due to the presence of suspended solid particles and the complexity of the fermentation matrix, while mixed substrate consumption and the presence of inhibitory compounds will further increase the complexity of the model. The choice of a suitable monitoring strategy depends on the model and the specific equipment requirements. Quantitative data (e.g., on accuracy, costs or concentration ranges) is desired for making objective decisions for control and optimization, but also to support and justify the choice of specific equipment. The collection of quantitative data is somewhat troublesome, as data from different sources either contradicted one another, as this could be dependent on the manufacturer and the specific reactor conditions, or was not available at all. The most reliable option, but also the most expensive and time-consuming one, is to test measurement equipment under practical conditions on a cellulose to ethanol fermentation plant and to make the results available to a broader public. It is not very realistic to assume that one organization can perform such tests alone. Therefore, it would be obvious to set up a consortium of stakeholders such that the test work – and the costs related to it – can be shared. It should also be in the interest of the measurement equipment manufacturers if an objective evaluation of the potential of the different measurement techniques would be available.

In Section 3 it was determined that the off-gas is the easiest to monitor in real-time because it avoids the interferences with the suspended solid particles. Also, off-gas analyzers that can detect oxygen, carbon dioxide or ethanol are often available in the industry. With this information, the controller can increase or decrease the batch times, and adjust the feeding rate based on the productivity of the fermentation. In section 5.1 it was evaluated that magnetic mass spectrometers are the most advantageous because they can evaluate a broad range of substances in a wide range of concentrations. Although the off-gas can give insight into the reactor characteristics, the evaluation of several models showed that the gas components such as carbon dioxide are hardly considered, while the ethanol stripping is not considered at all. Modeling the carbon dioxide concentration could potentially be useful in detecting uncommon behaviors in the system, as a deviation from mass balances might indicate that something is wrong in the process. However, it was shown that most models mainly consider substrates, products, biomass, and inhibitors, which can only be monitored in the liquid phase, and predictions based on off-gas only would not be as accurate. Contaminations by lactic acid bacteria can also be potentially monitored through mass balances and kinetics, but this option has not been thoroughly explored yet.

Monitoring the compounds dissolved in the liquid phase allows measuring the concentration of substrates, products, and inhibitors directly, giving a more clear picture of the actual state of the system. This information permits a better estimation of the biomass concentration and a control of the fermentation time and the feeding rate based on the actual concentrations of substrates and inhibitors. The main challenges are the interference with the suspended solid particles and the complex fermentation matrix of cellulose-to-ethanol fermentations. In this context, the choice of a monitoring method for the compounds in the liquid phase is not obvious and becomes a trade-off between the quality of the measured data, the speed of the analysis and the ease of the operation. On the one side of the spectrum, HPLC (score 12) is an excellent and well-known analytical tool with very high sensitivity and accuracy, but somewhat slow and complex to use. In addition to measuring substrates, products, and inhibitors, HPLC can measure the concentration of lactic acid, allowing the direct detection of contaminations by the LAB. On the other side of the spectrum, different in-line spectroscopies are easy to implement and have a high measuring frequency, but the measurements are noisy and less accurate. The accuracy of the spectroscopic methods improves when a filtration unit is added before the analysis, but this also increases the complexity of the operation. Among the different spectroscopic methods, in-line ATM-MIR is evaluated with the highest score (total score of 13) because it can measure the substrates, products, and lactic acid and it has been tested in demonstration scale cellulosic-ethanol fermentation. UV-Vis spectroscopy (score of 11) is also an interesting option as a fast on-line method to measure the concentration of inhibitors in the inlet or in the reactor. Biosensors (score of 10) obtained the lowest score, as they are sensitive and accurate methods to measure with high frequency the concentrations of glucose, xylose, ethanol or lactic acid, but they cannot be implemented in-line and require clear and diluted samples. The main challenges are that the sensors have limited long-term stability and will encounter drift, while there are also no furfural or 5-HMF biosensors available yet.

For biomass monitoring (section 5.3) dielectric spectroscopy was the most beneficial (total score of 12, Table 6) since it can differentiate cells from other suspended solid particles, it is able to detect viable cells, and it has been shown to work on lab-scale in cultures with lignocellulosic material. Although contaminations cannot be detected with this method, this study has shown other indirect methods to detect contaminations, such as the observation of a sudden increase in base addition to indicate lactic acid production from lactic acid bacteria. Unlike dielectric spectroscopy, flow cytometry (score of 8) can directly detect contaminations by lactic acid bacteria. However, flow cytometry is an expensive technique difficult to implement for on/at-line monitoring. 2D fluorescence and bio-calorimetry (scores of 7 and 8 respectively) are indirect methods to measure biomass, but they cannot detect contaminations. Finally, microscopy and image analysis (score of 11) appears as a method with the potential to measure biomass since it can differentiate cells from particles, viable and non-viable cells and also contamination. However, this method still needs further development.

When deciding on extending the monitoring scheme, one should first gain insight into what strategies will be the most useful for control and optimization. This will depend on how the process is modeled, but also on the type of process and the specific conditions applied. Off-gas measurements by mass spectrometry were found to be the most important in cellulosic ethanol fermentation, followed by the addition of the monitoring of the inlet. If it is assumed that the inlet composition is not dynamic, a delay in measurements is not an issue at all. HPLC is therefore suitable and reliable to monitor the inlet under this assumption. These two measurement techniques combined with kinetic models can generate data needed for control. Monitoring dissolved components and biomass in the reactor is of importance for fault detection and optimization, as this will need accurate data on the state of the reactor. A simulation study [89] including the addition of in situ measurements to estimate state variables, showed that the prediction error decreased when the reactor holdup, substrates, product, and pH were monitored with a sampling interval of 240 minutes. Interestingly, when excluding the pH from these measurements, the prediction error increased. Although total inorganic carbon was a state variable in this study, no off-gas monitoring was performed. It is recommended that a similar study is performed when a monitoring scheme is considered, to give a better insight into the added value of a specific monitoring scheme linked with a specific model.

Considering that cellulosic ethanol production processes have now reached a stage of maturity which allows operating a process at demonstration scale or even full-scale, it would be obvious to allocate some more resources to investigating the potential of further improving the operation of such installations by adding more on-line monitoring and control. In order to reach a situation where real-time control is put in operation on the basis of on-line measured data, our suggestion is, therefore, to focus on a detailed evaluation of the most promising monitoring methods that have been highlighted in this manuscript. As mentioned before, an *in-silico* approach could be useful here, inspired by the work on benchmarking of control strategies that has been done in the wastewater field [110].

# 9 Conclusion

Cellulose to ethanol fermentation is a complex process that is often operated far from its optimal conditions. In consequence, the implementation of advanced monitoring and control strategies is necessary to improve the process efficiency compared to non-cellulosic ethanol production processes.

Lignocellulosic waste includes a wide variety of materials ranging from wood chips to different kinds of straw. These materials have very different properties and compositions, and affect the fermentation differently. Likewise, the influence of the available process alternatives must be carefully considered before deciding on the most adequate monitoring and control system. In this review, different monitoring schemes and methods for cellulosic ethanol fermentation have been reviewed. The fermentation of wheat straw hydrolysate in an SHF process was used as a case study. However, the challenges described for this case study (e.g., high concentration of suspended solids, the complex fermentation matrix or the presence of inhibitors) are common to other substrates or process configurations.

The risk of contamination by lactic acid bacteria, the inhibition by furfural and acetic acid and the presence of oxygen in the fermenter were identified as the major threats for the cellulose to ethanol fermentation. Among the different monitoring schemes reviewed in this article, it was found that monitoring the off-gas, the inlet, and the liquid phase of the reactor would add significant value to the currently used monitoring methods (i.e., pH and temperature). Among all the methods available to monitor off-gas, only electronic noses and mass spectrometry are considered in this review as the two techniques able to simultaneously detect all the compounds of interest (glucose, xylose and ethanol). Despite the significant amount of research done in electronic noses, mass spectrometry is a more mature and implemented technology. To monitor the inlet and the liquid phase in the reactor, in-line ATR-MID spectroscopy was deemed as the most advantageous technique because it is able detect simultaneously most of the compounds of interest, it does not require sample preparation and it is not affected too much by the high concentrations of suspended solids. Monitoring the biomass was also found to be valuable. The most suited analytical instrument for real-time monitoring of the biomass is dielectric spectroscopy. However, the developments in microscopy and in image analysis make the technology attractive, especially for its potential to detect contaminations. It was found that quite some quantitative data on measuring devices is missing in the literature and that the available data can vary considerably depending on the manufacturer of a device, and on the reactor conditions. Research on the objective comparison of different devices in specific case studies or applications would be of interest, especially to companies aiming at selecting a device for a specific application.

Another important step is to investigate in more detail how the monitoring can contribute specifically to the control and optimization of industrial applications, and the most viable option there seems to use an in-silico approach to save on costs.

**Acknowledgments**

This work was partially financed by the European Regional Development Fund (ERDF) and Region Zealand (Denmark) through the BIOPRO-SMV project. Furthermore, the work received funding from Innovation Fund Denmark (BIOPRO2 strategic research center, project number 4105-00020B). We would like to acknowledge critical comments by Jesper Bryde-Jacobsen (BIOPRO), and Laila Thirup, Michael Elleskov, Pia Jørgensen, Flemming Mathiesen and Remus Mihail Prunescu from Ørsted. This project has also been supported partially by the EUDP project 'Demonstration of 2G ethanol in full scale, MEC' (Jr. no. 64015-0642). Finally, we wish to acknowledge the support obtained from the European Union's Horizon 2020 research and innovation programme under the Marie Sklodowska-Curie grant agreement number 713683 (COFUNDfellowsDTU) and from the Danish Council for Independent Research in the frame of the DFF FTP research project GREENLOGIC (grant agreement number 7017-00175A). Miguel Mauricio-Iglesias belongs to the Galician Competitive Research Group GRC2013-032 and the CRETUS strategic partnership (AGRUP2015/02), co-funded by FEDER (EU).

## 11 Supplementary material

### 11.1 Scoring method for the evaluation of the discussed methods to monitor the dissolved components

All methods were evaluated based on the following eight criteria: measured compounds, sensitivity, accuracy, drift, calibration and data analysis, sample preparation, response time, industrial implementation and costs.

The scores for measured compounds were based on the capabilities of each method to monitor key compounds of the cellulose to ethanol fermentation (Table S 1). A method capable of monitoring all the relevant compounds would receive a score of 3, whilst a method able to monitor none of the compounds would receive a score of 0. Methods able to monitor glucose

Table S 1. Scores given based on the capabilities to measure relevant compounds in the liquid phase.

|  | On-line NIR | In-line NIR | On-line MIR | In-line MIR | On-line Raman | In-line Raman | On-line UV-Vis | Biosensors* | At-line HPLC |
|---|---|---|---|---|---|---|---|---|---|
| Glucose | Yes | Yes | Yes | Yes | Yes | Yes | No | Yes | Yes |
| Xylose | Yes | Yes | Yes | Yes | Yes | Yes | No | Yes | Yes |
| Ethanol | Yes | Yes | Yes | Yes | Yes | Yes | No | Yes | Yes |
| Acetic acid | Yes | Yes | Yes | Yes | Yes | Yes | Yes | Yes | Yes |
| Lactic acid | No | No | Yes | Yes | Yes | Yes | Yes | Yes | Yes |
| Furfural | No | No | No | No | No | No | Yes | No | Yes |
| HMF | No | No | No | No | No | No | Yes | No | Yes |
| **Total score** | **1** | **1** | **2** | **2** | **2** | **2** | **1** | **2** | **3** |

Accuracy and sensitivity are evaluated based on the values found in the literature and discussed in Section 5 (Table S 2).

**Table S 2. Scores given based on the sensitivity and accuracy of each method.**

|  | On-line NIR | In-line NIR | On-line MIR | In-line MIR | On-line Raman | In-line Raman | On-line UV-Vis | Biosensors | At-line HPLC |
|---|---|---|---|---|---|---|---|---|---|
| **Sensitivity** | 1 | 1 | 2 | 2 | 2 | 2 | 1 | 2 | 3 |
| **Accuracy** | 2 | 1 | 2 | 1 | 2 | 1 | 2 | 2 | 3 |

Drift is evaluated based on the deviation of the measurements over time. All methods start with a maximum score of 3. Long-term deviations result in the subtraction of 1 point. Drift between and within batches results in the subtraction of 1 and 2 points, respectively (Table S 3).

**Table S 3. Scores given based on the basis of information collected about drift of each method.**

|  | On-line NIR | In-line NIR | On-line MIR | In-line MIR | On-line Raman | In-line Raman | On-line UV-Vis | Biosensors | At-line HPLC |
|---|---|---|---|---|---|---|---|---|---|
| Within batch | No | No | No | No | No | No | No | Yes | No |
| Between batches | Yes | Yes | Yes | Yes | No | No | No | No | No |
| Long-term deviations | Yes | Yes | Yes | Yes | Yes | Yes | Yes | Yes | Yes |
| **Drift** | 1 | 1 | 1 | 1 | 2 | 2 | 2 | 0 | 2 |

Scores related to the calibration and data analysis are based on two criteria: the complexity of calibration methods and the pre-processing requirements of each type of data. Univariate methods are the simplest ones and receive a score of 3, multivariate methods receive a score of 2 and multiway methods a score of 1. Preprocessing requirements are classified into P1 (including basic pre-processing techniques such as base-line correction or mean centering) and P2 (including P1 and additional methods to correct for other disturbances). A method requiring a pre-processing of type P1 or P2 would receive -1 or -2 points in their final scores, respectively (Table S 4).

**Table S 4. Scores given to each method according to the required calibration methods and data analysis. P1 includes basic pre-processing techniques such as base-line correction or mean centering. P2 includes P1 and additional methods to correct for other disturbances.**

|  | On-line NIR | In-line NIR | On-line MIR | In-line MIR | On-line Raman | In-line Raman | On-line UV-Vis | Biosensors | At-line HPLC |
|---|---|---|---|---|---|---|---|---|---|
| Univariate | No | No | No | No | No | No | No | Yes | Yes |
| Multivariate | Yes | Yes | Yes | Yes | Yes | Yes | Yes | No | No |
| Multiway | No | No | No | No | No | No | No | No | No |
| Pre-process | P1 | P2 | P1 | P2 | P1 | P2 | P1 | P1 | P1 |
| **Total score** | **1** | **0** | **1** | **0** | **1** | **0** | **1** | **2** | **2** |

The sample preparation is evaluated based on the number of steps required prior to analysis. A method requiring no sample preparation (in-line methods) would receive a score of 3, whilst methods requiring 1, 2, or 3 steps, would receive a score of 2, 1 or 0, respectively (Table S 5).

**Table S 5. Scores assigned to each method according to the sampling preparation requirements.**

|  | On-line NIR | In-line NIR | On-line MIR | In-line MIR | On-line Raman | In-line Raman | On-line UV-Vis | Biosensors | At-line HPLC |
|---|---|---|---|---|---|---|---|---|---|
| Filtration | Yes | No | Yes | No | Yes | No | Yes | Yes | Yes |
| Dilution | No | No | No | No | No | No | No | Yes | Yes |
| Derivation[1] | No | No | No | No | No | No | No | No | Yes |
| **Total score** | **2** | **3** | **2** | **3** | **2** | **3** | **2** | **1** | **0** |

[1] Derivation may include sample staining, or

The sampling frequency is divided into methods able to deliver almost real-time information (< 5 min), which receive a score of 3, methods with a delay of less than one hour (receiving a score between 2 if they need less than 20 minutes and 1 if they need more) and methods with a delay greater than one hour (receiving a score of 0) (Table S 6)

**Table S 6. Scores given to each method according to sample frequency.**

|  | On-line NIR | In-line NIR | On-line MIR | In-line MIR | On-line Raman | In-line Raman | On-line UV-Vis | Biosensors | At-line HPLC |
|---|---|---|---|---|---|---|---|---|---|
| < 5 min | No | Yes | No | Yes | No | Yes | No | No | No |
| < 1 hour | Yes | No | Yes | No | Yes | No | Yes | Yes | No |
| > 1 hour | No | No | No | No | No | No | No | No | Yes |
| **Total score** | **2** | **3** | **2** | **3** | **2** | **3** | **2** | **1** | **0** |

The evaluation of industrial implementation has been based on an extensive review of papers and patents. Industrial implementation refers to any fermentation process and it is not limited to cellulose to ethanol fermentations. Methods not implemented at industrial scale or that are rarely used would receive 0 and 1 point respectively, and methods commonly used at industrial scale would receive 2 points. Methods tested in large scale cellulose to ethanol fermentations would receive an additional point (Table S 7).

The scores regarding costs are divided into operational and investment costs and they are compared relatively to each other. A score of -3 is given to the most expensive equipment and a score of 0 is given to the cheapest one. The final score results from the rounded up average between the operational and the investment costs (Table S 8).

**Table S 7. Industrial implementation.**

|  | On-line NIR | In-line NIR | On-line MIR | In-line MIR | On-line Raman | In-line Raman | On-line UV-Vis | Biosensors | At-line HPLC |
|---|---|---|---|---|---|---|---|---|---|
| None | - | - | - | - | - | + | + | + | - |
| Rarely used | + | + | + | + | + | - | - | - | + |
| Commonly used | - | - | - | - | - | - | - | - | - |
| Tested in large scale 2G ethanol | + | + | + | + | - | - | - | - | - |
| **Total score** | **2** | **2** | **2** | **2** | **1** | **0** | **0** | **0** | **1** |

**Table S 8. Scores of each method related to the investment and operation costs.**

|  | On-line NIR | In-line NIR | On-line MIR | In-line MIR | On-line Raman | In-line Raman | On-line UV-Vis | Biosensors | At-line HPLC |
|---|---|---|---|---|---|---|---|---|---|
| Operation | -1 | 0 | -1 | 0 | -1 | 0 | -1 | -1 | -2 |
| Investment | -2 | -2 | -3 | -3 | -3 | -3 | -1 | 0 | -2 |
| **Total score** | **-1** | **-1** | **-2** | **-1** | **-2** | **-1** | **-1** | **0** | **-2** |

## 11.2 Scoring method used to evaluate the discussed methods to monitor biomass according to the different evaluation criteria

Each method is given 3 points if they are able to detect the corresponding feature (cells/particles, viable/dead or contaminations. The final score is obtained from the sum of each individual score.

**Table S 9. Scores based on the capabilities to differentiate cells and solid particles, to assess the viability of the cell culture and to detect contaminations.**

|  | OD probes | IR Spectroscopy | 2D fluorescence | Bio-calorimetry | Flow cytometry | Dielectric spectroscopy | Image analysis |
|---|---|---|---|---|---|---|---|
| **Cells/particles** | 0 | 0 | 3 | 3 | 3 | 3 | 3 |
| **Viable/dead** | 0 | 0 | 0 | 0 | 3 | 3 | 3 |
| **Contaminations** | 0 | 0 | 0 | 0 | 3 | 0 | 3 |
| **Total score** | **0** | **0** | **3** | **3** | **9** | **6** | **9** |

The sample preparation is evaluated based on the number of steps required prior to the analysis. A method requiring no sample preparation (in-line methods) would receive a score of 3, whilst methods requiring dilution, derivation or both, will receive between 0 and 2 points (Table S 10).

**Table S 10. Scores given to each method according to the sample preparation requirements.**

| | OD probes | IR Spectroscopy | 2D fluorescence | Bio-calorimetry | Flow cytometry | Dielectric spectroscopy | Image analysis |
|---|---|---|---|---|---|---|---|
| Dilution | Yes | No | No | No | Yes | No | Yes |
| Derivation | No | No | No | No | Yes | No | No |
| **Total score** | **2** | **3** | **3** | **3** | **0** | **3** | **2** |

Scores related to the calibration and data analysis are based on two criteria: the complexity of calibration methods and the pre-processing requirements of each type of data. Univariate methods are the simplest ones and receive a score of 3, multivariate receive a score of 2 and multiway methods and non-linear machine learning a score of 1. Preprocessing requirements are classified into P1 (including basic pre-processing techniques such as base-line correction or mean centering) and P2 (including P1 and additional methods to correct for other disturbances). A method requiring a pre-processing of type P1 or P2 would receive -1 or -2 points in their final scores, respectively (Table S 11).

**Table S 11. Scores given to each method according to the calibration and data analysis requirements. P1 includes basic pre-processing techniques such as base-line correction or mean centering. P2 includes P1 and additional methods to correct for other disturbances.**

| | OD probes | IR Spectroscopy | 2D fluorescence | Bio-calorimetry | Flow cytometry | Dielectric spectroscopy | Image analysis |
|---|---|---|---|---|---|---|---|
| Univariate | Yes | No | No | Yes | Yes | No | No |
| Multivariate | No | Yes | No | No | No | Yes | No |
| Multiway/Non-linear machine learning | No | No | Yes | No | No | No | Yes |
| Pretreatment | No | P1 | No | No | P1 | P1 | P1 |
| **Total score** | **3** | **1** | **1** | **3** | **2** | **1** | **0** |

The evaluation of industrial implementation has been based on an extensive review of papers and patents. Industrial implementation refers to any fermentation process and it is not limited to cellulose to ethanol fermentations. Methods not implemented at industrial scale or that are rarely used would receive 0 and 1 point respectively, and methods commonly used at industrial scale would receive 2 points. Methods tested in large scale cellulose to ethanol fermentations would receive an additional point (Table S 12).

The scores regarding costs are divided into operational and investment costs and they are compared relative to each other. A score of -3 is given to the most expensive equipment and a score of 0 is given to the cheapest one. The final score results from the rounded up average between the operational and the investment costs (Table S 13).

**Table S 12. Scores given to each method according to the industrial availability.**

|  | OD probes | IR Spectroscopy | 2D fluorescence | Bio-calorimetry | Flow cytometry | Dielectric spectroscopy | Image analysis |
|---|---|---|---|---|---|---|---|
| None | - | - | - | + | + | - | - |
| Rarely used | - | + | + | - | - | + | + |
| Commonly used | + | - | - | - | - | - | - |
| Tested in large scale cellulose-to ethanol | - | - | - | - | - | + | - |
| **Total score** | **2** | **1** | **1** | **0** | **0** | **2** | **1** |

**Table S 13. Scores given to each method according to operational and investment costs.**

|  | OD probes | IR Spectroscopy | 2D fluorescence | Bio-calorimetry | Flow cytometry | Dielectric spectroscopy | Image analysis |
|---|---|---|---|---|---|---|---|
| Operation | 0 | 0 | 0 | 0 | -3 | 0 | -1 |
| Investment | -1 | -2 | -2 | -2 | -3 | -1 | -1 |
| **Total score** | **0** | **-1** | **-1** | **-1** | **-3** | **0** | **-1** |